\documentclass[%reprint,
preprint,tightenlines,
nofootinbib,
 amsmath,amssymb,
 aps,
]{revtex4-1}

\usepackage{color,graphicx}% Include figure files
\usepackage{dcolumn}% Align table columns on decimal point
\usepackage{bm}% bold math
\newcounter{comment}

\newcommand{\red}{\textcolor{red}}
\begin{document}

\title{Measuring Polarized Gluon Distributions by Heavy Quark Spin Correlations and Polarizations}  

\author{Gary R.~Goldstein} 
\email{gary.goldstein@tufts.edu}
\affiliation{Department of Physics and Astronomy, Tufts University, Medford, MA 02155 USA.}

\author{Simonetta Liuti} 
\email{sl4y@virginia.edu}
\affiliation{Department of Physics, University of Virginia, Charlottesville, VA 22904, USA.}


 \begin{abstract}
The production of heavy flavor quark pairs, including top-anti-top, at the LHC proceeds primarily through gluon fusion. The correlation between the gluon spins affects various spin correlations between the produced quark and anti-quark. Both single spin asymmetries and double correlations of the quark pair spins will be manifest in the subsequent hadronization and decay distributions. For top pairs this is most pronounced. Dilepton, single lepton, and purely hadronic top pair decay channels allow for the extraction of gluon spin information as well as providing a window into possible interactions Beyond the Standard Model. Spin related asymmetries and polarizations will be presented. The implications for experimental determination will be discussed. 
\vspace{0.5in}

Talk presented at the APS Division of Particles and Fields Meeting (DPF 2017), July 31-August 4, 2017, Fermilab. C170731
\end{abstract}
\maketitle

\newcommand{\be}{\begin{equation}}
\newcommand{\ee}{\end{equation}}  

\section{Introduction}
In the following we will first review the 
%Parton Distribution Functions (pdf's), well known to high energy physicists, as extended to TMD's, and
Generalized Parton Distribution Functions (GPD's), that expand the phase space covered by Parton Distribution Function (pdf's) variables and relate to experimental processes involving exclusive leptoproduction of photons or hadrons. We will discuss a particular model for GPD's - the Reggeized spectator model referred to as the ``flexible parameterization" scheme.
For valence quarks, the exclusive processes connect to the underlying GPD's, and in Deeply Virtual Compton Scattering (DVCS), the gluon distributions contribute as well. Gluons will be our particular focus. We will show how the model predicts gluon GPDs, both unpolarized and polarized. The model, with parameterization fixed by various constraints, will be related to gluon ``transversity" and the associated Transverse Momentum Distribution (TMD) $h_1^g(x,\vec{k}_T^2)$. 
 
The production and decay of top-antitop pairs in hadron accelerators can, in principle, provide a measure of the gluon distributions, including the polarization. We will develop that interesting connection between gluon distributions, with and without polarization, and top-antitop spin correlations. 

%\noindent{\bf FORMALISM} \hspace{0.2cm} 
\section{Parton Distributions - quarks}
Deeply Virtual Compton Scattering  and Deeply Virtual Meson Production  can be described within QCD factorization, through the convolution of specific GPDs and hard scattering amplitudes.  There are four chiral-even GPDs, $H, E, \widetilde{H}, \widetilde{E}$ \cite{Ji_even} and
four additional chiral-odd GPDs, known to exist by considering twist-two quark operators that flip quark helicity by one unit, 
$H_T, E_T, \widetilde{H}_T, \widetilde{E}_T$ \cite{Ji_odd,Diehl_odd}.  
All GPDs depend on $(x,\xi,t, Q^2)$, two additional kinematical invariants besides the parton's  Light Cone (LC) momentum fraction, $x$, and the DVCS process' four-momentum transfer, $Q^2$,
namely $t=\Delta^2$ where $\Delta=P-P'$ is the momentum transfer between the initial and final protons, and $\xi$, or the fraction of LC momentum transfer, $\xi=\Delta^+/(P^+ + P'^+)$. 
%(Figure \ref{fig_handbag}). 
The {\it observables} containing the various GPDs are the Compton Form Factors (CFFs) - convolutions over $x$ of GPDs with the struck quark propagator. 
The quark GPDs are defined (at leading twist) as the matrix elements of  the following projection of the unintegrated quark-quark proton correlator (see Ref.\cite{Diehl_odd} for a detailed overview),
%\footnote{In what follows we can omit the Wilson gauge link without loss of generality.}  
\begin{eqnarray}
%{\cal M}_{i \, j}^{\Lambda \Lambda^\prime} (k,P,\Delta)=\int  d^4 y  \, {e^{iky}}\langle P^\prime, \Lambda^\prime | \overline{\psi}_{j}(0)\psi_{i}(y) | P,  \Lambda \rangle,
W_{\Lambda', \Lambda}^\Gamma(x,\xi,t) & = & \frac{1}{2} \int \frac{d z^- }{2 \pi} e^{ix\overline{P}^+ z^-} \left. \langle P', \Lambda' \mid \overline{\psi}\left(-\frac{z}{2}\right) \Gamma \, \psi\left(\frac{z}{2}\right)\mid P, \Lambda \rangle \right|_{z^+=0,{\bf z}_T=0},
\label{matrix}
\end{eqnarray}
where $\Gamma=\gamma^+, \gamma^+\gamma_5, i\sigma^{i+}\gamma_5 (i=1,2)$, and the target's spins are $\Lambda, \Lambda^\prime$. 
For the two chiral-even cases
\begin{eqnarray}
%\label{correlator1}
W_{\Lambda', \Lambda}^{[\gamma^+]}(x,\xi,t)  & = &  \frac{1}{2\overline{P}^+} \overline{U}(P',\Lambda') \left( \gamma^{+} H(x,\xi,t) +
 \frac{i\sigma^{+\mu}(- \Delta_\mu)}{2M}  E(x,\xi,t)   \right)U(P,\Lambda); \label{correlator1} \\
W_{\Lambda', \Lambda}^{[\gamma^+\gamma^5]}(x,\xi,t)  & = &  \frac{1}{2\overline{P}^+} \overline{U}(P',\Lambda') \left( \gamma^{+} \gamma^5 {\tilde H}(x,\xi,t) +
\gamma^5 \frac{- \Delta^+}{2M}  {\tilde E}(x,\xi,t)   \right)U(P,\Lambda) 
\label{correlator2}
\end{eqnarray}

% \nonumber \\
%& + & \left.  \frac{P^+ \Delta^i - \Delta^+ P^i}{M^2}  \widetilde{H}_T(x,\xi,t)  +
%\frac{\gamma^+ P^i - P^+ \gamma^i}{2M} \widetilde{E}_T(x,\xi,t) \right) U(P,\Lambda)
%\end{eqnarray}

For the chiral-odd case, $\Gamma= i\sigma^{i+}\gamma_5$, $W_{\Lambda', \Lambda}^\Gamma$ was parametrized as \cite{Diehl_odd},
\begin{eqnarray}
\label{correlator}
W_{\Lambda', \Lambda}^{[i\sigma^{i+}\gamma_5]}(x,\xi,t)  & = &  \frac{1}{2\overline{P}^+} \overline{U}(P',\Lambda') \left( i \sigma^{+i} H_T(x,\xi,t) +
 \frac{\gamma^+ \Delta^i - \Delta^+ \gamma^i}{2M} E_T(x,\xi,t)   \right. \nonumber \\
& + & \left.  \frac{P^+ \Delta^i - \Delta^+ P^i}{M^2}  \widetilde{H}_T(x,\xi,t)  +
\frac{\gamma^+ P^i - P^+ \gamma^i}{2M} \widetilde{E}_T(x,\xi,t) \right) U(P,\Lambda)
\end{eqnarray}

The quark chiral even GPDs connect to pdf's in the forward limit
\be
H^q(x, 0, 0) = q(x)=h^q_1(x) \;\; {\rm and} \;\; \widetilde{H}^q(x,0,0) = \Delta q(x) = q(x)_\Rightarrow^\rightarrow - q(x)_\Rightarrow^\leftarrow =g^q_1(x)
\ee
and the chiral even GPDs integrate to the nucleon  form factors, which constrains the GPD t-dependence,  
%\begin{subequations}
\begin{equation}
\int_0^1 H^q(X,\zeta,t) = F_1^q(t),  \int_0^1 E^q(X,\zeta,t) = F_2^q(t),
\int_0^1 \widetilde{H}^q(X,\zeta,t) = G_A^q(t), \int_0^1 \widetilde{E}^q(X,\zeta,t) = G_P^q(t). 
\label{GP}
\end{equation}
%\end{subequations}
where $F_1^q(t)$ and $F_2^q(t)$ are the Dirac and Pauli form factors for the quark $q$ components in the nucleon. $G_A^q(t)$ and $G_P^q(t)$ are the axial  and pseudoscalar form factors. 
%Furthermore,  $H(x,0,0)=h_1(x)$ and $\widetilde{H}(x,0,0)=g_1(x)$. 

The GPDs can be connected with a 3-dimensional picture of the constituents within the proton through the Fourier Transform over $\Delta$ to impact parameter space at fixed values of x (or $\xi$)~\cite{Burkardt}. They also allow the decomposition into quark and gluon angular momenta - spin and orbital angular momenta. There have been several theoretical models for GPDs and the phenomenological applications to experimental data of some of those models have been successful . The quark GPDs are constrained by well measured pdf's and EM form factors, but the gluon GPDs have less explicit connection to measured quantities. One model for the quark GPDs, the ``flexible model", has successfully parameterized measurements of DVCS as well as $\pi^0$ electroproduction. That model is a spectator picture with the nucleon Fock states dominated by a quark and a spectator diquark.  

The spin structures of GPDs that are directly related to spin dependent observables are most effectively expressed in term of helicity dependent amplitudes, developed extensively for the covariant description of two body scattering processes 
(see also Ref.\cite{Diehl_odd}).

\section{Gluon GPDs}
The helicity conserving gluon distributions with t-channel even parity are defined : ($\bar{P}^+ = \frac{P^+ + P^{\prime+}}{2}$)
\begin{eqnarray}\label{eqn:1}
F^g &=& \frac{1}{\bar{P}^+}\int \frac{dz^-}{2\pi}e^{ix\bar{P}^+z^-}\langle P^{\prime},\Lambda^{\prime}|G^{+\mu}(-\frac{1}{2}z)G_{\mu}{}^{+}(\frac{1}{2}z) |P,\Lambda\rangle \Big|_{z^+=0,\vec{z}_T=0} 
%\rightarrow \hspace{120pt} \nonumber\\
%\frac{1}{\bar{P}^+}\int \frac{dz^-}{2\pi}e^{ix\bar{P}^+z^-}\langle P^{\prime},\Lambda^{\prime}|G^{+j}(-\frac{1}{2}z)G^{+j}(\frac{1}{2}z) |P,\Lambda \rangle \Big|_{z^+=0,\vec{z}_T=0} = \hspace{120pt}
 \nonumber\\
&=& \frac{1}{2\bar{P}^+}\bar{U}(P^{\prime},\Lambda^\prime)[H^g(x,\xi,t) \gamma^+ + E^g(x,\xi,t) \frac{i\sigma^{+\alpha}(-\Delta_{\alpha})}{2M}]U(P,\Lambda) 
%\hspace{100pt} 
\end{eqnarray}
and for t-channel odd parity  
\begin{eqnarray}\label{eqn:2}
\tilde{F^g} &=& \frac{-i}{\bar{P}^+} \int \frac{dz^-}{2\pi} e^{ix\bar{P}^+z^-}\langle P^{\prime},\Lambda^{\prime}|G^{+\mu}(-\frac{1}{2}z)\tilde{G}_{\mu}{}^{+}(\frac{1}{2}z) |P,\Lambda\rangle \Big|_{z^+=0,\vec{z}_T=0} %\rightarrow \hspace{120pt} \nonumber \\
%\frac{-i}{\bar{P}^+} \int \frac{dz^-}{2\pi} e^{ix\bar{P}^+z^-}\langle P^{\prime},\Lambda^{\prime}|G^{+j}(-\frac{1}{2}z)\tilde{G}^{+j}(\frac{1}{2}z) |P,\Lambda \rangle\Big|_{z^+=0,\vec{z}_T=0} = \hspace{120pt} 
\nonumber \\
&=&\frac{1}{2\bar{P}^+}\bar{U}(P^{\prime},\Lambda^\prime)[\tilde{H}^g(x,\xi,t) \gamma^+\gamma_5 + E^g(x,\xi,t) \frac{\gamma_5(-\Delta^+)}{2M}]U(P,\Lambda), 
%\hspace{100pt} 
\end{eqnarray}
summing over transverse indices $j=1,2$, and using the dual gluon field strength
$\tilde{G}^{\mu\nu}(x) = \frac{1}{2} \epsilon^{\mu\nu\alpha\beta}G_{\alpha\beta}(x)$.\footnote{With this convention  $H_g$ reduces to the pdf $xg(x,0,0)$.}
The transverse polarization components enter here because we are considering leading order (twist 2) 
%and the ``good" components of the gluon polarization 4-vectors are transverse polarization (helicity $\pm 1$)- 
on-shell (in the light cone quantization method, Ref.\cite{Diehl_odd}).
The longitudinal polarization (helicity 0) enters at twist 3. There are also  contributions involving the transverse helicity flip ($\pm 1 \rightarrow \mp 1$), which can be thought of as gluon states of transversity, or equivalently, linear polarization states. 

The gluon ``transversity" distributions are defined as (Ref.~\cite{Diehl_odd})
\begin{eqnarray}\label{eqn:1}
&& F_T^g = -\frac{1}{\bar{P}^+}\int \frac{dz^-}{2\pi}e^{ix\bar{P}^+z^-}\langle P^{\prime},\Lambda^{\prime}|{\bf S} G^{+j}(-\frac{1}{2}z)G^{+ k}(\frac{1}{2}z) |P,\Lambda\rangle \Big|_{z^+=0,\vec{z}_T=0} %\rightarrow \hspace{120pt} 
\nonumber\\
&=& {\bf S} \frac{1}{2 \bar{P}^+} \frac {\bar{P}^+ \Delta^j - \Delta^{+} \bar{P}^j }{2M\bar{P}^+} \nonumber \\
&&%\int \frac{dz^-}{2\pi}e^{ix\bar{P}^+z^-}\langle P^{\prime},\Lambda^{\prime}|G^{+i}(-\frac{1}{2}z)G^{+i}(\frac{1}{2}z) |P,\Lambda \rangle \Big|_{z^+=0,\vec{z}_T=0} = \hspace{120pt} \nonumber\\
\times %\frac{1}{2\bar{P}^+}
\bar{U}(P^{\prime},\Lambda^\prime)\left[ H_T^g(x,\xi,t) i\sigma^{+k} +{\tilde H}_T^g\frac {\bar{P}^+ \Delta^k - \Delta^{+} \bar{P}^k }{M^2} \right. \nonumber \\
&& \hspace{1.5in}\left. + E_T^g(x,\xi,t) \frac{\gamma^+ \Delta^k - \Delta^+ \gamma^k}{2M} +{\tilde E}_T^g \frac{\gamma^+ \bar{P}^k - \bar{P}^+ \gamma^k}{M}  \right] U(P,\Lambda) 
%\hspace{100pt} 
\end{eqnarray}
wherein ${\bf S}$ symmetrizes in $(j,k)$ and removes the trace ~\cite{Diehl_odd}. 

The double helicity flip {\bf does not mix} with quark distributions, which makes gluon transversity unique and useful. In the definition of transversity~\cite{GG-MJM} for on-shell gluons or photons, wherein there are no helicity 0 states, the transversity states are
\begin{eqnarray}
\mid +1)_{trans} &=& \{ \mid +1 \rangle + \mid -1 \rangle \} /2 = \mid -1 )_{trans} \nonumber \\
\mid 0)_{trans} &=& \{ \mid +1 \rangle - \mid -1 \rangle \} /\sqrt{2} \nonumber \\
{\rm helicity} \quad \mid \pm 1 \rangle &=& \{ \mp \hat{x} - i\hat{y} \}/\sqrt{2} \nonumber \\
\hat{x} &=& -\mid 0 )_{trans} = P_{parallel} \nonumber \\
\hat{y} &=& i \sqrt{2} \mid +1 )_{trans} = P_{normal}
\end{eqnarray}
where the two-body scattering plane is the X-Z plane, with $\hat{y}$ along the normal to the scattering plane.

Our approach to modeling and parameterizing the valence quark GPDs~\cite{GGL}
has a natural generalization to the gluon and sea quark GPDs~\cite{GLgluons}. The key ingredients for the valence quarks are the spectator model and the Reggeization.  
To begin with, in our spectator model for gluon GPDs the nucleon decomposes into a gluon and a color octet baryon, so that the overall color is a singlet (projected from the $8\otimes 8 = 1\oplus 8 \oplus 8^\prime \oplus 10 \oplus 27$). The color octet baryon has the same flavor as the nucleon and is a Fermion (with color$\otimes$flavor$\otimes$spin being antisymmetric under quark label exchanges), which we take to be spin 1/2 for simplicity. This can be realized with the 70 representation of the flavor-spin SU(6), containing flavor SU(3) $\otimes$ spin SU(2) representations (8,1/2) $\oplus$ (10, 1/2) $\oplus$ (8, 3/2) $\oplus$ (1,1/2). Of these, only (8, 1/2) and (8, 3/2) contain Isospin = 1/2 states with nucleon flavor. To keep the model simple, we need only take the (8,1/2) as the spectator. The overall $8_{{\rm color}} \otimes 70_{{\rm flavor-spin}}$ must be antisymmetrized under exchange of any pair of quarks, resulting in a particular combination of the flavor-spin (8,1/2) and (8,3/2). Taking only the spin 1/2, though, provides sufficient parameterization to fit the $H_g(x,0,0)$ to the pdf g(x). Once the gluon distribution is given transverse momentum, through $t$, and skewness, via $\xi$, the spin 3/2 spectator can contribute to double gluon helicity flip as readily as the spin 1/2 spectator.

Evolving with $Q^2$ also requires a sea quark contribution, which we take in a spectator picture with $N \rightarrow {\bar u} \oplus (uuud)$ or ${\bar d} \oplus (uudd)$. They are also ``normalized" by fitting parameters to the phenomenologically determined sea quark pdf's. These contributions are of interest also, particularly in applications to exclusive neutrino photon production.

The helicity amplitudes,  $A^{g \, *}_{\Lambda^{\prime},\Lambda_{g^\prime};\Lambda,\Lambda_g}$,   %combinations are 
can be expressed in terms of the GPDs (Ref.~\cite{Diehl_odd}). For the gluon helicity conserving amplitudes, 
\begin{eqnarray}
A_{++,++} &=& \sqrt{1-\xi^2}\Big(\frac{H^g+\tilde{H}^g}{2}-\frac{\xi^2}{1-\xi^2}\frac{E^g+\tilde{E}^g}{2}\Big)\nonumber \\
A_{-+,-+} &=& \sqrt{1-\xi^2}\Big(\frac{H^g-\tilde{H}^g}{2}-\frac{\xi^2}{1-\xi^2}\frac{E^g-\tilde{E}^g}{2}\Big)\nonumber \\
A_{++,-+} &=& -e^{-i\phi}\frac{\sqrt{t_0-t}}{2M}\Big(\frac{E^g-\xi\tilde{E}^g}{2}\Big)\nonumber \\
A_{-+,++} &=& e^{i\phi}\frac{\sqrt{t_0-t}}{2M}\Big(\frac{E^g+\xi\tilde{E}^g}{2}\Big),
%\nonumber \\ \nonumber
\end{eqnarray}
where $\phi$ is the azimuthal phase angle of 
\begin{equation}
D=\frac{p^\prime}{1-\xi}-\frac{p}{1+\xi}
\end{equation} 
($\vec{D} = \frac{\vec{P^{\prime}_{\perp}}}{1-\xi} = \frac{-\vec{\Delta_{\perp}}}{1-\xi}$ for frames in which $\vec{P}_{\perp} = 0$ and $e^{i\phi} = e^{i\phi_{\Delta}+\pi}=-(\Delta_1 + i\Delta_2)/\mid {\vec \Delta}_T \mid $. Also  $\sqrt{t_0-t} = \frac{|\Delta_{\perp}|}{\sqrt{1-\zeta}}$. ) 

For the gluon double helicity flip amplitudes,
\begin{eqnarray}
A_{++,+-} &=& e^{2i\phi}\sqrt{1-\xi^2}\, \frac{t_0 -t}{4M^2} \Big({\tilde H}_T^g+(1-\xi)\frac{E_T^g+\tilde{E}_T^g}{2}\Big)\nonumber \\
A_{-+,--} &=& e^{2i\phi}\sqrt{1-\xi^2}\, \frac{t_0 -t}{4M^2} \Big({\tilde H}_T^g+(1+\xi)\frac{E_T^g-\tilde{E}_T^g}{2}\Big)\nonumber \\
%A_{-+,-+} &=& \sqrt{1-\xi^2}\Big(\frac{H^g-\tilde{H}^g}{2}-\frac{\xi^2}{1-\xi^2}\frac{E^g-\tilde{E}^g}{2}\Big)\nonumber \\
A_{++,--} &=& +e^{i\phi} (1-\xi^2) \frac{\sqrt{t_0-t}}{2M} \Big( H_T^g +\frac{t_0-t}{M^2} \tilde{H}_T^g - \frac{\xi^2}{1-\xi^2} E_T^g +  \frac{\xi}{1-\xi^2} \,\tilde{E}_T^g\Big)\nonumber \\
A_{-+,+-} &=& - e^{3i\phi} (1-\xi^2) \frac{\sqrt{t_0-t}^3}{8M^3}\tilde{H}_T^g,
%\nonumber \\ \nonumber
\label{gluonflip}
\end{eqnarray}
similar to the quark helicity flip amplitudes.
%\footnote{The phases and signs are in agreement with Ref.~\cite{Diehl_hab}, wherein each helicity flip amplitude for quarks is multiplied here by the complex conjugate of that reference's overall factor $e^{+i\phi} \sqrt{1-\xi^2}\sqrt{t_0-t}/2M$.}

In a forthcoming publication~\cite{GLgluons} we will present our explicit model for the gluon GPDs, generalizing from the Regge-diquark spectator model, the ``flexible model". We will address some questions that are unique to gluon distributions: how are the {\it t} and skewness $\xi$ dependences normalized? How is the small x behavior accounted for? What is the connection to the Pomeron? 

In brief, the spectator model will take direct, point-like ``vertex functions" $\mathcal{G}_{\Lambda_X; \Lambda_g, \Lambda}(x,\vec{k}_T^2)$ for $N(\Lambda) \rightarrow g(\Lambda_g) + X(\Lambda_X)$ and their conjugate to construct the s-channel and u-channel gluon+nucleon helicity amplitudes,  $A^{g \, *}_{\Lambda^{\prime},\Lambda_{g^\prime};\Lambda,\Lambda_g}$. One simplification that results from this spectator picture is that
\be
\widetilde{H}_T^g = 0, \; (1-X) A_{- +, - -} = (1 -X') A_{+ +, + -},  \;  \widetilde{E}_T^g = 0,
\ee
as in ref.~\cite{ Hood-Ji}. The model for the gluon GPDs answers the questions raised above regarding observables for different processes. 

To begin, in exclusive electroproduction of photons via high virtuality photon exchange (i.e. at large $Q^2$),  or DVCS, the model restricts gluon distributions. 
\begin{figure}
\includegraphics[width=15cm]{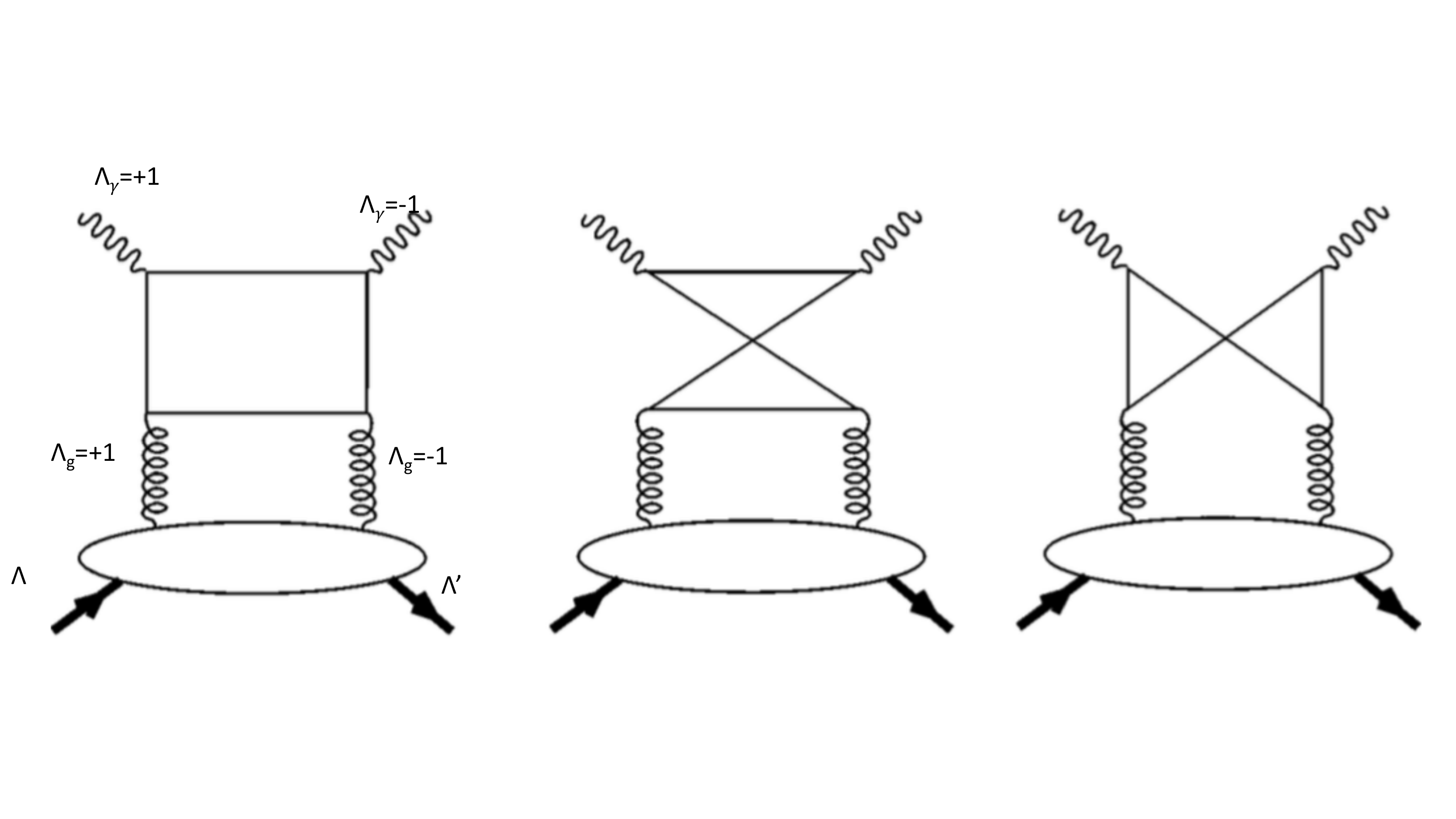}
\caption{Order $\alpha_S$ quark loop calculation for $\gamma+N$ via gluon distributions. See Ref.~\cite{Hood-Ji}}
\label{photon-gluon}
\end{figure}
To measure the transversity GPD's of the gluons in DVCS or DVMP (with neutral vector mesons), electroproduction requires order $\alpha_s$ quark loop amplitudes (see Fig.~\ref{photon-gluon}). These appear as azimuthal modulations of the differential cross section at 4th order in sines and cosines. The gluon transversity GPDs can be separated from the helicity conserving gluon GPDs in the interference cross section for Bethe-Heitler and DVCS phase modulations. The multiply differential cross section for DVCS is
\begin{equation}
\frac{d^5\sigma}{d x_{Bj} d Q^2 d|t| d\phi d\phi_S } =
\frac{\alpha^3}{16\pi^2 (s-M^2)^2 \sqrt{1+\gamma^2}}
\big|T\big|^2 \;,
\label{eq:xs5fold}
\end{equation}
where $\alpha$ is the electromagnetic fine structure constant,
$\gamma = 2 M x_{Bj}/Q$, $s=(k+p)^2$, and $M$ is the mass of the target,
and $T$ is a
coherent superposition of the DVCS and Bethe-Heitler amplitudes,
\begin{equation}
T(k, p, k^\prime, q^\prime, p^\prime) = T_{DVCS}(k, p, k^\prime, q^\prime, p^\prime) + T_{BH}(k, p, k^\prime, q^\prime, p^\prime),
\end{equation}
%
%
%can be described,  at order $\alpha_{EM}^2$, by the helicity amplitudes,
yielding,
\begin{equation}
|T|^2 = |T_{\rm BH} + T_{\rm DVCS}|^2
=|T_{\rm BH}|^2 + |T_{\rm DVCS}|^2 + \mathcal{I}\;.
\label{eq:coherent}
\end{equation}
%for Deeply Virtual Compton Scattering (DVCS) and the Bethe-Heitler (BH), respectively.
\begin{eqnarray}
\label{interf}
\mathcal{I} & = & T_{BH}^{*} T_{DVCS}
+ T_{DVCS}^{*} T_{BH} .
\end{eqnarray}
%%%%%%%%%%%%%%%%%%%%% FIGURE KINEMATICS

The transversity gluons appear in the $\cos(3\phi)$ modulation due to the double flip of the gluon helicity combined with the single flip of the BH amplitude. The product of the amplitudes gives rise to a term in the cross section modulation (involving the convolution of the GPD with the quark loop that couples to the photons)~\cite{Diehl_odd, Diehl-etal,BMK}
\begin{equation}
\frac{\sqrt{t_0-t}^3}{8M^3}\left[H^g_T F_2 - E^g_T F_1 - 2{\tilde H}^g_T \left(F_1+\frac{t}{4M^2} F_2\right)\right] \cos3\phi .
\end{equation}

In hadronic collision processes, the gluon distributions are folded into the more probable initial and final state interactions. Nevertheless, we will see that at the LHC, the production of top pairs can enhance the 
ability to separate out a form of polarized gluon contributions.

\section{Top-antitop Spin Correlations}
Before the discovery of the top quark at the Fermilab Tevatron, one proposed method to disentangle the signal for top quark production from the daunting background of multiple hadron events was to concentrate on the spin correlations of the top and antitop decay products. The ``golden events'' were expected to be the dilepton events in which two very energetic opposite sign leptons would signal the weak decays of each top into b-quarks and W's, the latter decaying leptonically. At the energies accessed by the Tevatron, the primary mechanism for production of the top-antitop pair was correctly expected to be light quark-antiquark annihilation. The actual observations of top quarks by the D0 and CDF groups did not use the spin correlations. Nevertheless, these correlations provide a test of the QCD mechanism and a version of those correlations was roughly confirmed by D0~\cite{d0spin,parke}.

The LHC now produces many more top quarks, but the higher energy makes quark-antiquark annihilation less important than another mechanism, gluon fusion. Gluon fusion, involving the merging of two vector particles, gives rise to quite distinct spin correlations among the top decay products. In this paper we will present the spin density matrices and angular correlations for both mechanisms. 

\begin{figure}
\includegraphics[width=18cm]{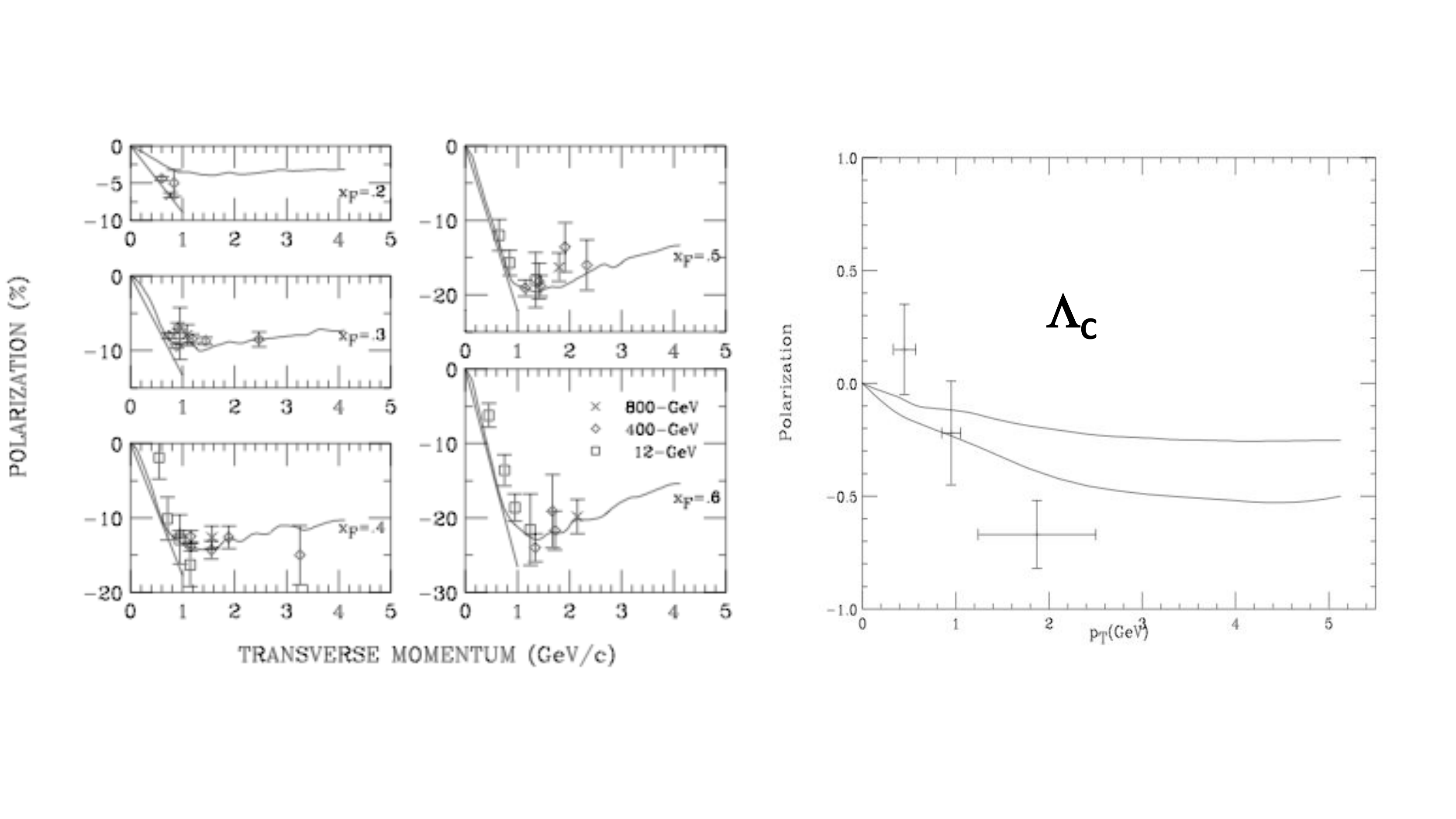}
\caption{Left: $\Lambda$ polarization (~\cite{Heller}) with curves from Dharmaratna and Goldstein, Ref.~\cite{DharmaGoldst}. Right: Prediction based on extending the model to the charm sector ~\cite{GGLambda_c} with data from Ref.~\cite{Ramberg}.}
\label{hyperonSSA}
\end{figure}

\begin{figure}
\includegraphics[width=18cm]{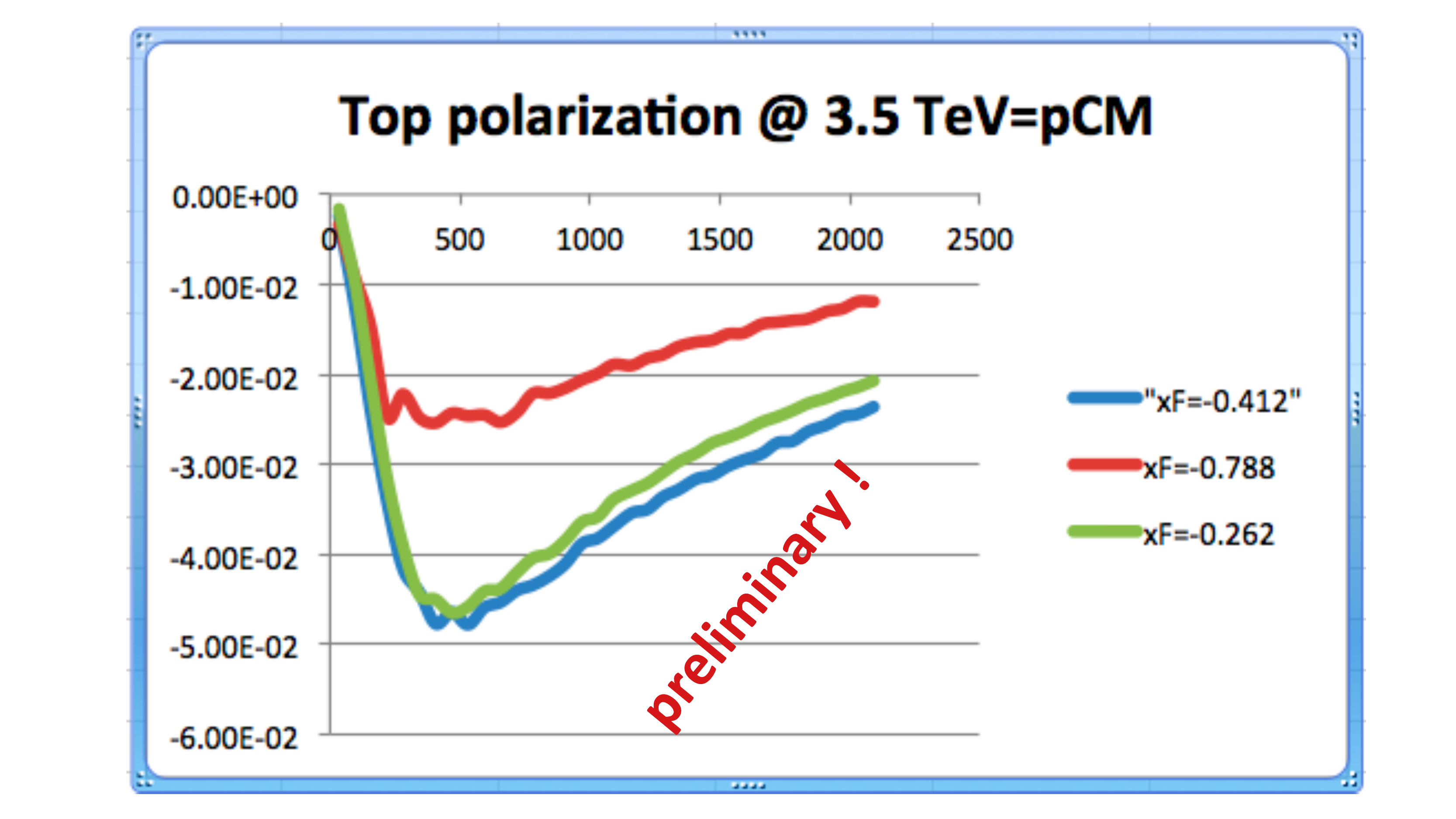}
\caption{Predicted top polarization from p+p collisions at 3.5 TeV vs. $p_T$ at 3 values of $x_F$, based on perturbative QCD model of  Dharmaratna and Goldstein Ref.~\cite{DharmaGoldst}.}
\label{t-pol}
\end{figure}

%Those correlations will be sensitive to the top quark mass and, with high enough statistics, could further constrain that mass. At this time there remains some differences in mass determinations from the Tevatron in different channels. Some examples of how the spin correlations can narrow the mass distributions will be shown.
%%%%%%%%%%%%% SSA for TOP%%%%%%
First we can ask what is known about single top or antitop polarization? This is an important question, because it bears on hadron scattering, large single spin asymmetry (SSA) measurements of baryons - strange and charmed. Recent determinations of top SSA at the LHC are small - from ATLAS $A_p=-0.035\pm0.040$~\cite{ATLAS} and from CMS $A_p=0.005\pm0.01$~\cite{CMS}. Large SSA's for strange hyperons were first observed in hadron collisions in 1976~\cite{Bunce, Heller}, contrary to expectations from Perturbative QCD~\cite{KPR}. An explanation based on one loop QCD calculations with an ansatz for ``recombination"~\cite{DharmaGoldst}, was able to fit data on $\Lambda$ and $\Lambda_c$, as shown in Fig.~\ref{hyperonSSA}, with collinear quark pdf's. Using the QCD calculation, extended to top quarks, and with a simple form for the gluon distributions - the collinear pdf's - sizable polarization results. The polarization peaks close to $-0.05$ vs. $p_T$, over a range of $x_F$ as shown in Fig.~\ref{t-pol}.

This prediction is within the small values and their uncertainties determined by both CMS and ATLAS.

\vspace{0.4cm}
The spin correlations for the {\bf top-antitop pairs} produced by unpolarized $p+\bar{p}$ or $p+p$ collisions can be calculated precisely at tree level QCD for the quarks or gluons and folded into the relevant parton distribution functions. At this point there are enough top pairs in the data from ATLAS and CMS to begin to us the spin correlations as probes of the production mechanism. There has always been considerable interest in the distribution of gluons in the nucleon. It has become clear that heavy hadron production~\cite{BoeBro} or even Higgs boson production~\cite{Boer} at very high energies could provide a measure of the {\it polarized gluon} distributions in the protons. It is now possible, and especially interesting to use the top pair spin correlation as a lever to disentangle the gluon polarization distributions.
%\footnote{to quote V. Telegdi, ``Yesterday's sensation is today's calibration"}.

In the following we present the tree level production mechanisms for $g_{X\, {\rm or}\, Y}\, + g_{X\, {\rm or}\, Y}\, \rightarrow t_{\pm}\, + {\bar t}_{\pm}\, + X$, where the gluon subscripts are linear polarization directions and the t-quark subscripts are helicities. We will rely heavily on helicity amplitudes. We will see that, due to parity conservation in the production process and the fact that the gluons are not directly observed, the top pair spin observables can select the linear polarization of the gluons more directly than the helicity of the gluons.

Considerable work has been done in predicting and subsequently measuring top spin correlations at the LHC since the above predictions were made. A standard parametrization is to represent the top-antitop cross section asymmetries as
\begin{equation}
\frac{1}{\sigma}  \frac{d^2\sigma}{d\cos\theta_1 d\cos\theta_2}=\frac{1}{4}(1 +B_1\cos\theta_1 +B_2 \cos\theta_2 -C_{helicity} \cos\theta_1 \cdot \cos\theta_2 )
\label{LHC-param}
\end{equation}
where the polar angles $\theta_1, \, \theta_2$, for the decay product leptons from the top and antitop, are measured relative to the $t$-direction in the $t+\bar{t}$ center of momentum. Such a distribution corresponds to experimentally summing over all other kinematic variables for the leptons. The measurements of $C_{helicity}$ by ATLAS and CMS agree with the QCD calculations of Ref.~\cite{Bernreuther}.
In general, however, there are azimuthal dependences as well. These will be important for different polarization initial states. We will develop the full angular dependences of the top-antitop spin correlations as they depend on gluon distributions.

\section{Gluon-top pair observables}
The QCD amplitudes for $g+g\rightarrow t+ {\bar t}$ are well known at tree level~\cite{DharmaGoldst,parke}. There are 3 Feynman diagrams and two distinct color couplings. With on-shell gluons there are 2 helicity values, $\pm 1$, leading to 16 hellcity combinations, but parity restricts the number of independent amplitudes to 8. Let these amplitudes be $A_{\Lambda_{g1}, \Lambda_{g2} ; \,t, \, {\bar t}}({\hat s},{\hat t})$ with $\Lambda_{g1}, \Lambda_{g2}$ the gluon helicities,  $t, {\bar t}$ the top and antitop helicities, and ${\hat s},{\hat t}$ the kinematic invariants in an arbitrary frame. Let $g^{(1)}_{\Lambda_{N1}, \Lambda_{X1}, \Lambda_{g1}}(x_1,k_T,\,M_{X1}^2)$ be the amplitude for proton number 1 to emit a gluon no.1 with longitudinal momentum fraction $x_1$ and transverse momentum $k_T$, along with an unspecified residual $X1$ of mass $M_{X1}$, with corresponding $g^{(2)}$ for the other proton. The overall amplitude for $N_1 + N_2 \rightarrow t+ X_1 +{\bar t} + X_2$ is then
\begin{equation}
%\sum \int_{X_1, \,X_2} 
g^{(1)}_{\Lambda_{N1}, \Lambda_{X1}, \Lambda_{g1}}g^{(2)}_{\Lambda_{N2}, \Lambda_{X2}, \Lambda_{g2}}  A_{\Lambda_{g1}, \Lambda_{g2} ; t, {\bar t}}
\label{ttbar_amp}
\end{equation}
The kinematic variables of the gluon from proton number 1 in $g^{(1)}$ must be matched with the hard amplitude $A$ and the other incoming gluon $g^{(2)}$. Here they are implicit and will be discussed later.

\begin{figure}
\includegraphics[width=12cm]{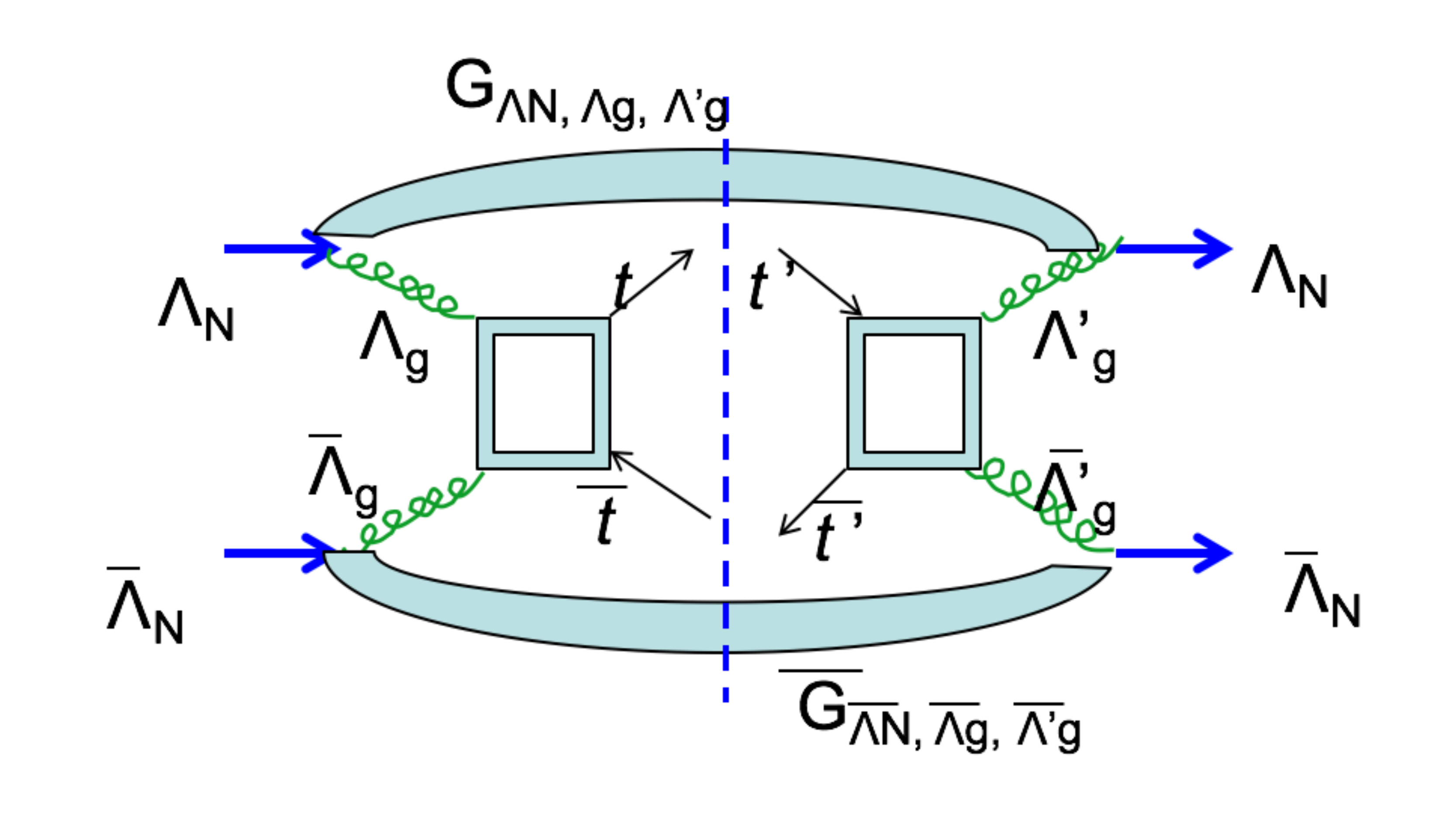}
\caption{Illustration of cross section for definite helicities in $p+p \rightarrow t+\bar{t} +X$ }
\label{gluon-ttbar}
\end{figure}

To construct differential cross sections for unpolarized colliding protons, this amplitude must be combined with its conjugate and summed and integrated over unobserved quantities,
\begin{eqnarray}
\sum \int_ {X_1,X_2} g^{(1)}_{\Lambda_{N1}, \Lambda_{X1}, \Lambda_{g1}}g^{(2)}_{\Lambda_{N2}, \Lambda_{X2}, \Lambda_{g2}}  A_{\Lambda_{g1}, \Lambda_{g2} ; t, {\bar t}}g^{(1)*}_{\Lambda_{N1}, \Lambda_{X1}, \Lambda^\prime_{g1}}g^{(2)*}_{\Lambda_{N2}, \Lambda_{X2}, \Lambda^\prime_{g2}}  A^*_{\Lambda^\prime_{g1}, \Lambda^\prime_{g2} ; t^\prime, {\bar t}^\prime}
\label{ttbar_cross}
\end{eqnarray}
with the summation and integration over $X_1,X_2,\Lambda_{g1},\Lambda_{g2},\Lambda^\prime_{g1},\Lambda^\prime_{g2},\Lambda_{N1},\Lambda_{N2}$ - see Fig.~\ref{gluon-ttbar}. The terms can be rearranged to correspond to gluon distributions and hard scattering amplitude products,
\begin{eqnarray}
\sum_{\Lambda_{g1},\Lambda_{g2},\Lambda^\prime_{g1},\Lambda^\prime_{g2}} &&\left(\sum_{\Lambda_{N2},\Lambda_{X2}} \int_ {X_2}  g^{(2)*}_{\Lambda_{N2}, \Lambda_{X2}, \Lambda^\prime_{g2}} g^{(2)}_{\Lambda_{N2}, \Lambda_{X2}, \Lambda_{g2}}\right) \nonumber \\
&&\times \left(\sum_{\Lambda_{N1},\Lambda_{X1}} \int_ {X_1} g^{(1)*}_{\Lambda_{N1}, \Lambda_{X1}, \Lambda^\prime_{g1}} g^{(1)}_{\Lambda_{N1}, \Lambda_{X1}, \Lambda_{g1}} \right) 
 A^*_{\Lambda^\prime_{g1}, \Lambda^\prime_{g2} ; t^\prime, {\bar t}^\prime}A_{\Lambda_{g1}, \Lambda_{g2} ; t, {\bar t}}
\label{ttbar_cross}
\end{eqnarray}
as illustrated in Fig.~\ref{gluon-ttbar}.

The two bracketed terms are the gluon distributions, 
\begin{equation}
G^{(1)}_{\Lambda_{N1},\Lambda_{g1},\Lambda^\prime_{g1}}=\sum_{\Lambda_{X1}} \int_ {X_1} g^{(1)*}_{\Lambda_{N1}, \Lambda_{X1}, \Lambda^\prime_{g1}} g^{(1)}_{\Lambda_{N1}, \Lambda_{X1}, \Lambda_{g1}} ,
\label{G1}
\end{equation}
and similarly for $G^{(2)}$.
\begin{figure*}
%\centering
\includegraphics[width=8cm]{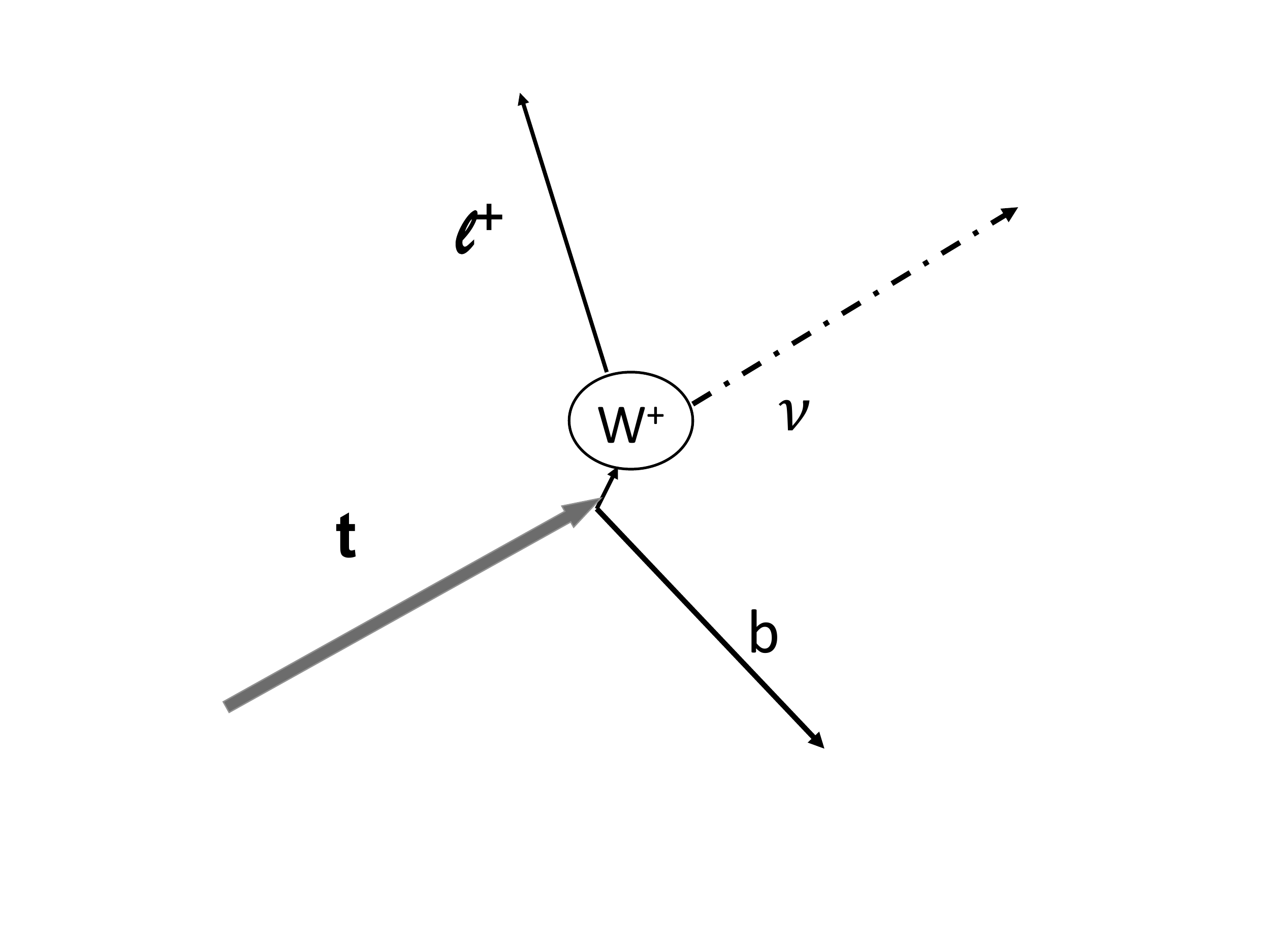}
\hspace{0.1mm}
\includegraphics[width=8cm]{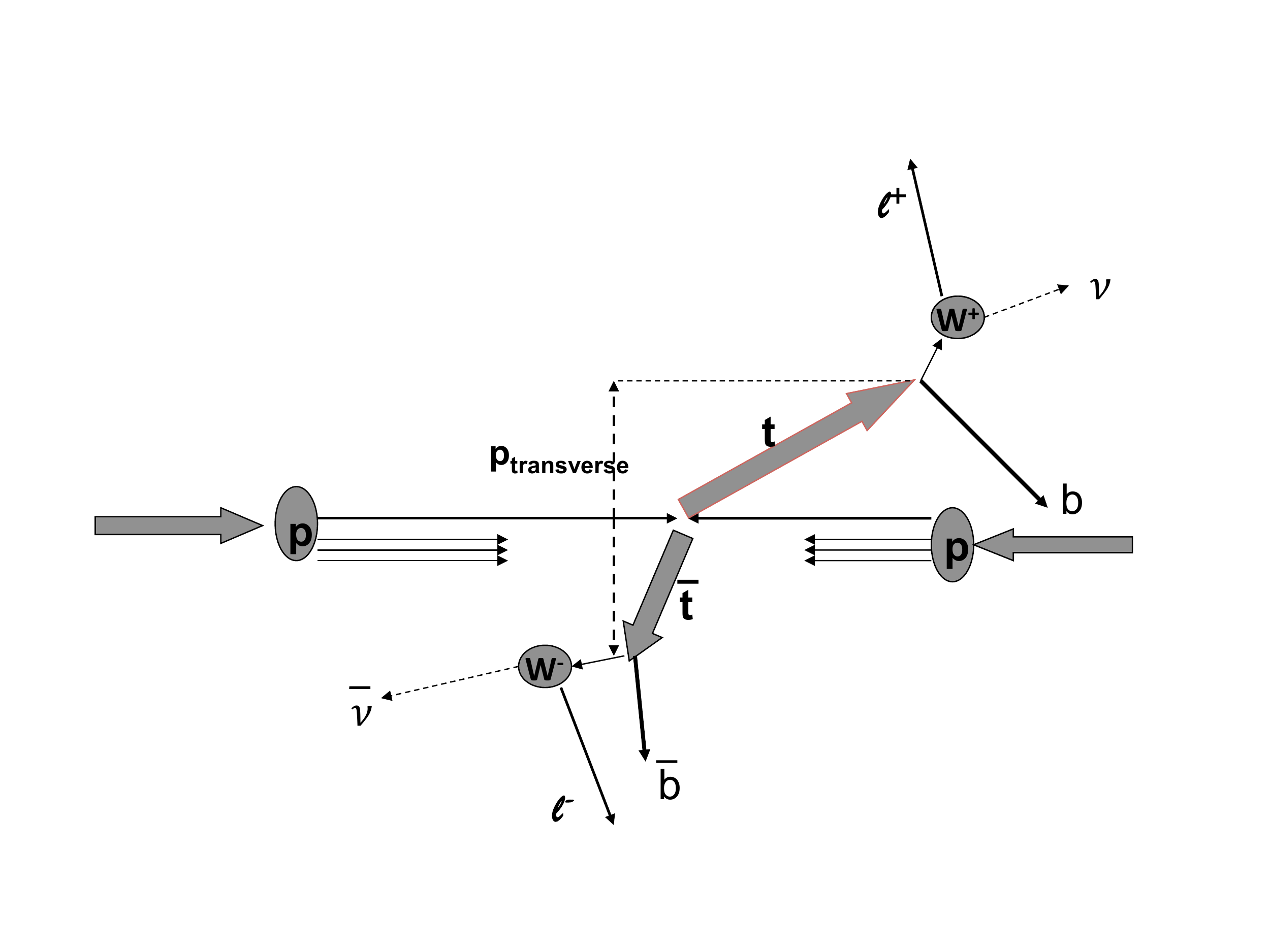}
\caption{Leptonic top decay channel (left), $t+\bar{t}$ production and dilepton double decay channel (right).}
%}
%%%%% IT LOOKS LIKE CANNOT USE \cite IN THE CAPTION - STRANGE?  %%%%%
%[\refcite{GGLeta}]. }
%Entries were collected in S. Liuti, INT Workshop, Feb. 2014. Circled prediction is from ref~
%\cite{hybrid_odd}.
%}
\label{t-decay}       % Give a unique label
\end{figure*}

Then the multiple sum can be written more compactly as a double density matrix,
\begin{equation}
\rho_{t^\prime,{\bar t}^\prime; t, {\bar t}}=  \sum_{\Lambda_{g1},\Lambda_{g2},\Lambda^\prime_{g1},\Lambda^\prime_{g2}} \sum_{\Lambda_{N2},\Lambda_{N1}} G^{(2)}_{\Lambda_{N2},\Lambda_{g2},\Lambda^\prime_{g2}}G^{(1)}_{\Lambda_{N1},\Lambda_{g1},\Lambda^\prime_{g1}} A^*_{\Lambda^\prime_{g1}, \Lambda^\prime_{g2} ; t^\prime, {\bar t}^\prime}A_{\Lambda_{g1}, \Lambda_{g2} ; t, {\bar t}}.
\label{G1G2}
\end{equation}

Note that parity conservation constrains the $G$'s and the $A$'s via
\begin{eqnarray}
G^{(1\,({\rm or} 2))}_{-\Lambda_{N1},-\Lambda_{g1},-\Lambda^\prime_{g1}} &=&G^{(1\,({\rm or} 2))}_{\Lambda_{N1},\Lambda_{g1},\Lambda^\prime_{g1}}
\nonumber \\
A_{-\Lambda_{g1},- \Lambda_{g2} ; -t, -{\bar t}} &=& (-1)^{t - {\bar t}}A_{\Lambda_{g1}, \Lambda_{g2} ; t, {\bar t}}
\label{parity}
\end{eqnarray}
 
 When the summations over the various unmeasured helicities are carried out and the parity relations are used, four distinct terms arise,
 %%% get structure now --- fill in right subscripts later
\begin{eqnarray}
\rho_{t^\prime,{\bar t}^\prime; t, {\bar t}}= \left[\left(G^{(2)}_{+,R,R} + G^{(2)}_{+,L,L}\right) \right. &&
\left[ \left[ A^*_{RR,t^\prime,{\bar t} }A_{RR,t,{\bar t}} + A^*_{RL,t^\prime,{\bar t} }A_{RL,t,{\bar t}} \right. \right. \nonumber \\
&&\left. \left. +A^*_{LR,t^\prime,{\bar t} }A_{LR,t,{\bar t}} + A^*_{LL,t^\prime,{\bar t}}A_{LL,t,{\bar t}} \right] \left(G^{(1)}_{+,R,R} + G^{(1)}_{+,L,L}\right) \right. \nonumber \\
&& +\left[A^*_{LR,t^\prime,{\bar t} }A_{RR,t,{\bar t}} + A^*_{LL,t^\prime,{\bar t} }A_{RL,t,{\bar t}} \right. \nonumber \\
&& \left. \left. +A^*_{RR,t^\prime,{\bar t} }A_{LR,t,{\bar t}} + A^*_{RL,t^\prime,{\bar t}}A_{LL,t,{\bar t}} \right] \left(G^{(1)}_{+,R,L} + G^{(1)}_{+,L,R}\right) \right. \nonumber \\
+\left(G^{(2)}_{+,R,L} + G^{(2)}_{+,L,R}\right) &&\left[ \left[ A^*_{RL,t^\prime,{\bar t} }A_{RR,t,{\bar t}} + A^*_{RR,t^\prime,{\bar t} }A_{RL,t,{\bar t}} \right. \right. \nonumber \\
&&\left. \left. +A^*_{LL,t^\prime,{\bar t} }A_{LR,t,{\bar t}} + A^*_{LR,t^\prime,{\bar t}}A_{LL,t,{\bar t}} \right] \left(G^{(1)}_{+,R,R} + G^{(1)}_{+,L,L}\right) \right. \nonumber \\
&& +\left[A^*_{LL,t^\prime,{\bar t} }A_{RR,t,{\bar t}} + A^*_{LR,t^\prime,{\bar t} }A_{RL,t,{\bar t}} \right. \nonumber \\
&& \left. \left. +A^*_{RL,t^\prime,{\bar t} }A_{LR,t,{\bar t}} + A^*_{RR,t^\prime,{\bar t}}A_{LL,t,{\bar t}} \right] \left(G^{(1)}_{+,R,L} + G^{(1)}_{+,L,R}\right) \right]
\label{bigsum}
\end{eqnarray}
The subscripts $R,L$ correspond to gluon helicities $\pm 1$. Because of the parity relations Eq.~\ref{parity}, the combination of gluon distributions that appear in the summation is limited. The two independent combinations correspond to linear polarization states,
\begin{eqnarray}
G^{(1)}_{\Lambda_{N1},R,R} + G^{(1)}_{\Lambda_{N1},L,L}&=& G^{(1)}_{\Lambda_{N1},XX} + G^{(1)}_{\Lambda_{N1},YY} = G^{(1)}_{\Lambda_{N1} ,\, UP} 
\label{UP} \\
G^{(1)}_{\Lambda_{N1},R,L} + G^{(1)}_{\Lambda_{N1},L,R}&=&G^{(1)}_{\Lambda_{N1},YY} - G^{(1)}_{\Lambda_{N1},XX} = G^{(1)}_{\Lambda_{N1} ,\, LP}
\label{LP}
\end{eqnarray} 
The $UP$ and $LP$ subscripts on the right are for unpolarized and linearly polarized gluons. The ${\hat X}\, \& \,{\hat Y}$ directions are transverse to the gluon 3-momentum ${\vec k}_1$, with ${\hat X}$ in the $g_1+g_2\rightarrow t+{\bar t}$ scattering plane. For gluon number 2, the 3-momentum ${\vec k}_2$
is neither parallel nor anti-parallel to ${\vec k}_1$, in general, but the $X-Z$-planes coincide. So the ${\hat X}$ direction for $g_2$ differs from $g_1$, but the ${\hat Y}$ directions coincide. Care must be taken with the helicity and linear polarization labels for $g_2$ and ${\bar t}$ because their 3-momenta are anti-parallel to the corresponding $g_1$ and $t$ 3-momenta in the $t+{\bar t}$ CM. It is easiest to visualize the subprocess in the CM frame. That frame is operationally determined from the observed $t\, {\bar t}$ pair in the laboratory, boosting back to their CM. The direction of the boost fixes a $Z$-axis from which a polar angle for the top in the CM can be determined. We will specify the kinematics more carefully below.

The cumbersome form of Eq.~\ref{bigsum} can be simplified using the notation of Eqs.~\ref{UP}, \ref{LP} and the definitions
\begin{subequations}
\label{rhos}
\begin{eqnarray}
\rho^{UP,UP}_{t^\prime,{\bar t}^\prime; t, {\bar t}}&=&\left[ A^*_{RR,t^\prime,{\bar t} }A_{RR,t,{\bar t}} + A^*_{RL,t^\prime,{\bar t} }A_{RL,t,{\bar t}} 
%\right.  \nonumber \\
%&&\left.  
+A^*_{LR,t^\prime,{\bar t} }A_{LR,t,{\bar t}} + A^*_{LL,t^\prime,{\bar t}}A_{LL,t,{\bar t}} \right] \\
\rho^{UP,LP}_{t^\prime,{\bar t}^\prime; t, {\bar t}}&=&\left[A^*_{LR,t^\prime,{\bar t} }A_{RR,t,{\bar t}} + A^*_{LL,t^\prime,{\bar t} }A_{RL,t,{\bar t}} 
%\right. \nonumber \\
%&& \left. \left. 
+A^*_{RR,t^\prime,{\bar t} }A_{LR,t,{\bar t}} + A^*_{RL,t^\prime,{\bar t}}A_{LL,t,{\bar t}} \right] \\
\rho^{LP,UP}_{t^\prime,{\bar t}^\prime; t, {\bar t}}&=&\left[ A^*_{RL,t^\prime,{\bar t} }A_{RR,t,{\bar t}} + A^*_{RR,t^\prime,{\bar t} }A_{RL,t,{\bar t}} 
%\right. 
%\right. \nonumber \\
%&&\left. \left. 
+A^*_{LL,t^\prime,{\bar t} }A_{LR,t,{\bar t}} + A^*_{LR,t^\prime,{\bar t}}A_{LL,t,{\bar t}} \right] \\
\rho^{LP,\red{LP}}_{t^\prime,{\bar t}^\prime; t, {\bar t}}&=&\left[A^*_{LL,t^\prime,{\bar t} }A_{RR,t,{\bar t}} + A^*_{LR,t^\prime,{\bar t} }A_{RL,t,{\bar t}} 
%\right. \nonumber \\
%&& \left. \left. 
+A^*_{RL,t^\prime,{\bar t} }A_{LR,t,{\bar t}} + A^*_{RR,t^\prime,{\bar t}}A_{LL,t,{\bar t}} \right] 
\end{eqnarray}
\end{subequations}
so that 
\begin{eqnarray}
\rho_{t^\prime,{\bar t}^\prime; t, {\bar t}}=\sum_{\Lambda_{N1},\Lambda_{N2}}&& \{G^{(2)}_{\Lambda_{N2} ,\, UP} \,\rho^{UP,UP}_{t^\prime,{\bar t}^\prime; t, {\bar t}}\,G^{(1)}_{\Lambda_{N1} ,\, UP} +G^{(2)}_{\Lambda_{N2} ,\, UP} \,\rho^{UP,LP}_{t^\prime,{\bar t}^\prime; t, {\bar t}}\,G^{(1)}_{\Lambda_{N1} ,\, LP} \nonumber \\
& +& G^{(2)}_{\Lambda_{N2} ,\, LP} \,\rho^{LP,UP}_{t^\prime,{\bar t}^\prime; t, {\bar t}}\,G^{(1)}_{\Lambda_{N1} ,\, UP} + G^{(2)}_{\Lambda_{N2} ,\, LP} \,\rho^{LP,LP}_{t^\prime,{\bar t}^\prime; t, {\bar t}}\,G^{(1)}_{\Lambda_{N1} ,\, LP}\}
\end{eqnarray}  
 
The next step is to evaluate the tree level hard scattering amplitudes $A_{\Lambda_{g1}, \Lambda_{g2} ; t, {\bar t}}$. These can be evaluated in the CM frame in terms of the variables ${\hat s}, \, \theta, \, \beta$ and the color factors for the $(8)\otimes(8)\rightarrow (3)\otimes ({\bar 3})$. The latter involve the $f_{a\,b\,c}$ and $d_{a\,b\,c}$ couplings, or more simply the combinations of Gell-Mann matrices $(\lambda^b\lambda^c)_{j\,k}$ and $(\lambda^c\lambda^b)_{j\,k}$ where $(a,b,c)$ are octet labels and $(j,\,k)$ are triplet-anti-triplet labels. Aside from overall normalization factors, each helicity amplitude $A_{\Lambda_{g1}, \Lambda_{g2} ; t, {\bar t}}$ can be written in the form
\begin{equation}
\label{color}
A_{\Lambda_{g1}, \Lambda_{g2} ; t, {\bar t}}|_{(j,\,k)}^{(b,\,c)}=\left[\frac{(\lambda^b\lambda^c)_{j\,k}}{(m_t^2-{\hat t})}a^t_{\Lambda_{g1}, \Lambda_{g2} ; t, {\bar t}} + \frac{(\lambda^c\lambda^b)_{j\,k}}{(m_t^2-{\hat u})}a^u_{\Lambda_{g1}, \Lambda_{g2} ; t, {\bar t}} \right]
\end{equation}

In Table~\ref{dtable} we show the values of the density matrix elements obtained from these amplitudes.
It is now apparent that different kinematic regions of spin correlated $t-{\bar t}$ pairs will select out different combinations of polarized gluon distributions.

In deriving the expressions in Table~\ref{dtable} we have stayed in the top-antitop CM and we let this be a reference plane, X-Z plane. But the top-antitop pair can be azimuthally dependent, relative to that plane, in which the gluon linear polarization is defined. To determine how the azimuthal dependence enters these density matrix elements, we first rotate the 2-spinors from z-axis quantization to the $\theta,\phi$ direction via 
\begin{equation}
\chi_\lambda(k,\theta, \phi) = e^{-i\sigma_z \frac{\phi}{2}}e^{-i\sigma_y \frac{\theta}{2}}e^{+i\sigma_z \frac{\phi}{2}} \chi_\lambda(k,0,0)
\end{equation}
Using this rotation we can see that only amplitudes with helicity pairs $\lambda_t = -\lambda_{{\bar t}}$ will
involve azimuthal factors $e^{\pm i \phi}$. This comes about because the 3 tree level Feynman diagrams for the $a^t_{\Lambda_{g1}, \Lambda_{g2} ; t, {\bar t}}$ (and 3 for the $a^u$) in Eq.~\ref{color} can be reduced to the 2-spinor form  
\begin{equation}
\chi_t^\dagger(k,\theta,\phi) {\mathcal M} \chi_{{\bar t}}(k,\pi-\theta,\pi+\phi),
\label{Mttbar}
\end{equation}
where ${\mathcal M}$ is a $2 \times 2$ matrix that consists of terms proportional to $(I, \sigma_z, \sigma^\pm=(\sigma_x\pm i\sigma_y)/\sqrt{2})$. Of these terms, only $\sigma^\pm$ pick up a phase under the rotation 
\begin{equation}
e^{-i\sigma_z \phi} \sigma^\pm e^{+i\sigma_z \phi}=\pm e^{\mp 2i \phi} \sigma^\pm .
\end{equation}
and $\sigma^\pm$ are raising and lowering operators for the antitop helicity spinor on the right in Eq.~\ref{Mttbar}. This makes sense, as follows. In order for the $t+{\bar t}$ pair to have an azimuthal dependence two conditions have to be met. The net gluon spin component along one gluon momentum ${\vec p}_1$ must be $\pm 2$ (for on-shell gluons) in order to establish a spin direction. Its total helicity cannot be 0. The same requirement applies to the top pair. That requires the pair to have opposite helicities. The spin component along ${\vec k}$ will be $\pm 1$. Then the amplitude will be of the form of $sin^n(\theta) e^{\pm 2 i \phi}$. The double density matrix elements will involve  azimuthal dependences $I, e^{\pm 2 i \phi}, e^{\pm 4 i \phi}$. Summing over all gluon helicities (i.e. unpolarized gluon distributions) is equivalent to adding the three columns in Table~\ref{dtable}, which cancels out the azimuthal dependence.
\begin{figure}
%\begin{center}
\includegraphics[width=15cm]{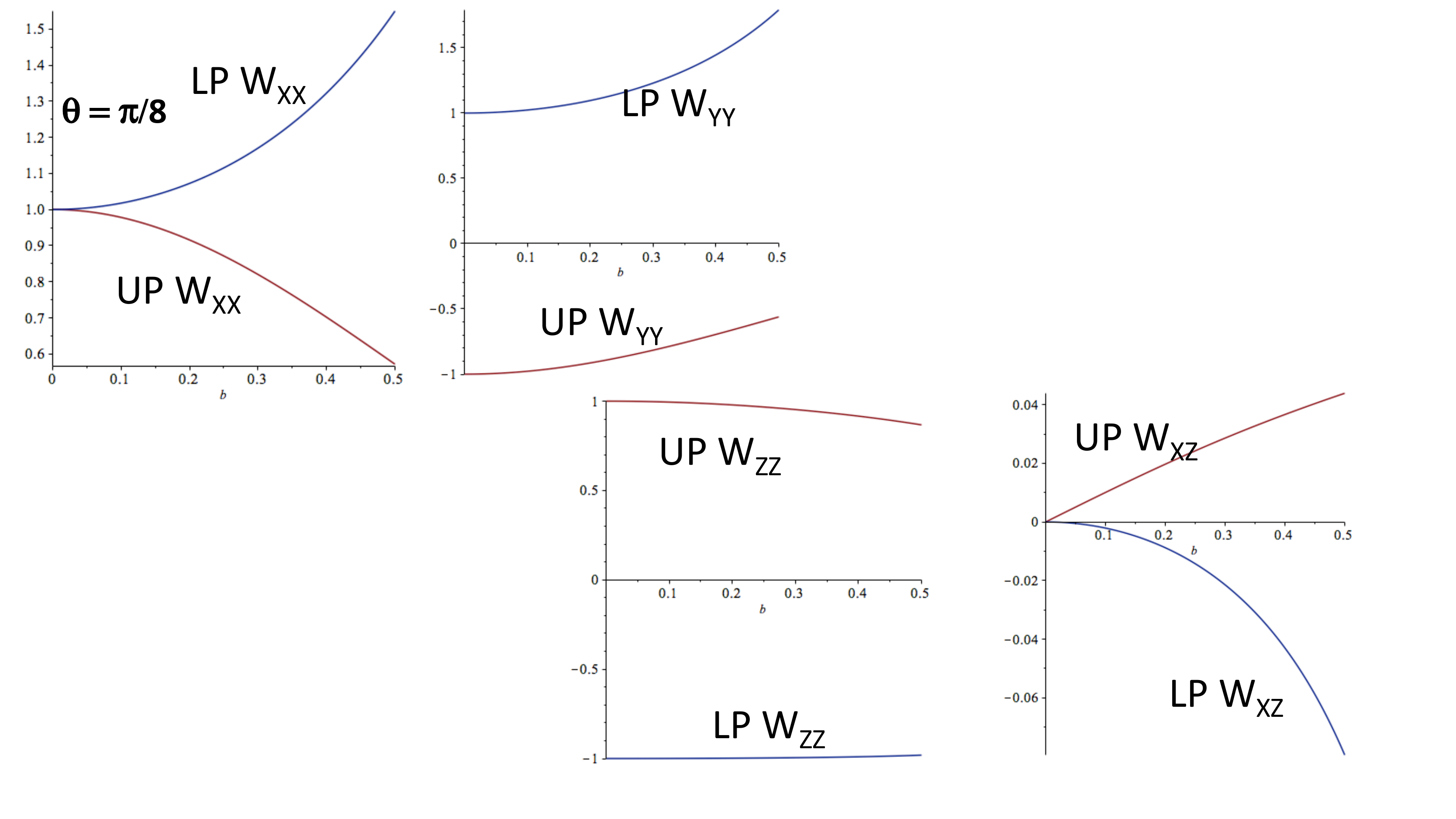}
%\end{center}
\caption{Weighting for Cartesian components of $\hat{p}(\mu^+), \; \hat{p}(\mu^-)$, plotted for {\bf varying $\beta$}, the magnitude of relativistic velocity of the top in the $t+\bar{t}$ Center-of-Mass frame. Each is plotted for unpolarized and transverse-linear polarized gluon distributions. Each lepton momentum is evaluated in the corresponding top or anti-top rest frame, with directions defined by CM.}
\label{weights}
\end{figure}
%%%%%%%%%% continue with azimuthal  

With this table and the relations to the gluon helicities or transversities, we are now able to incorporate top-antitop decays as markers of polarizations. We next complete the relation between top decay distributions and polarization correlations.
% {\bf MUCH MORE TO FILL IN - the next section has to be changed to be relevant}

\begin{table}
%\label{densitytable}
\begin{tabular}{|c|c|c|c|}
\hline
$\rho_{t^\prime,{\bar t}^\prime; t, {\bar t}}$  & UP,UP &LP,LP & UP,LP + LP,UP  \\
 \hline
 ++, ++ & $\gamma^{-2}(1+\beta^2(1+sin^4\theta))$ & $\gamma^{-2}(-1+\beta^2(1+sin^4\theta))$ & $-4\gamma^{-2}\beta^2 sin^2\theta$ \\
 \hline
$+ -, +-$   & $\beta^2 sin^2\theta ( 2 - sin^2\theta))$ & $-\beta^2sin^4\theta$ & 0  \\
 \hline
$ ++, - -$ & $\gamma^{-2}(-1+\beta^2(1+sin^4\theta))$ & $\gamma^{-2}(+1+\beta^2(1+sin^4\theta))$ & $+4\gamma^{-2}\beta^2 sin^2\theta$ \\
 \hline
$+ -, - +$   & $\beta^2sin^4\theta$&$-\beta^2 sin^2\theta ( 2 - sin^2\theta))$  & 0  \\ 
\hline
$+ +, + - $ & $-2\gamma^{-1} \beta^2 sin^3\theta cos\theta$ & $-2\gamma^{-1}\beta^2 sin^3\theta cos\theta $ & $-4\gamma^{-1}\beta^2 sin\theta cos\theta$ \\
\hline
$+ +, - + $ & $2\gamma^{-1} \beta^2 sin^3\theta cos\theta$ & $2\gamma^{-1}\beta^2 sin^3\theta cos\theta $ & $4\gamma^{-1}\beta^2 sin\theta cos\theta$ \\
\hline
\end{tabular}
\caption{Values, for gluon production of top pairs, of double density matrix elements $\rho$ for combinations in Eq.~\ref{rhos} using values of helicity amplitudes from Eq.~\ref{color} , evaluated in the $t+{\bar t}$ center of mass.}
\label{dtable}
\end{table}

\section{Top decay distributions}

The semi-leptonic decays of the top quark afford the best opportunity for
polarization analysis~\cite{DalGol1}. The opposite-sign leptons
usually have very high transverse momenta and are accompanied by b-quark
jets. So the double correlation of top spins are manifested in the
joint decay distributions into leptons and b-jets. The decay is primarily through the favored $t \rightarrow W^+ +b; \, W^+ \rightarrow l^+ + \nu_l$. As shown in Ref.~\cite{DalGol2}, the amplitude $B_{\lambda_b,t}$ for a polarized top quark at rest to decay into a measured b-quark and antilepton along with an unobserved neutrino has the simple angular dependence given by 
\begin{eqnarray}
U_{t,{\bar t}}&=&\sum_{\lambda_b} B_{\lambda_b,{\bar t}}^* B_{\lambda_b,t} \nonumber \\
&\propto& (I+{\vec p}_{\bar l} \cdot {\vec \sigma}_t/p_{\bar l})_{t, {\bar t}} (p_b \cdot p_\nu ),
\end{eqnarray}
with the spin dependence factorizing into this simple form.

The top spin correlations are expressed as double density matrix
elements.
%~\cite{dalitz3} (or as a direct product form~\cite{chen}). 
The quark spin correlations are transmitted to the decay
products, shown in Fig.~\ref{t-decay}. The correlations between the lepton directions and the parent
top spin (in the top rest frame) produce correlations between the lepton
directions, which has been expressed as a weighting factor~\cite{spin96},
The light quark-antiquark annihilation mechanism, for {\bf unpolarized} quarks, gives rise to the angular distribution
between opposite charge lepton pairs,  
\begin{eqnarray}
W(\theta,p,p_{\bar l},p_l) & = & \frac{1}{4}
\left\{1+[\sin^2\theta([p^2+m^2](\hat{p}_{\bar l})_x(\hat{p}_l)_{\bar{x}} +
[p^2-m^2] (\hat{p}_{\bar l})_y(\hat{p}_l)_{\bar{y}})\right. \nonumber\\
                           &   & \mbox{} - 2 m p \cos\theta
\sin\theta((\hat{p}_{\bar l})_x(\hat{p}_l)_{\bar{z}} + (\hat{p}_{\bar
l})_z(\hat{p}_l)_{\bar{x}})  + ([p^2-m^2]\nonumber\\
                           &   & \mbox{} \left. +[p^2+m^2] \cos^2\theta)
(\hat{p}_{\bar l})_z(\hat{p}_l)_{\bar{z}}]
            / [(p^2+m^2)+(p^2-m^2)cos^2\theta]\right\} \\
&=& \frac{1}{4} + \frac{1}{4} \left\{ (2-\beta^2) \sin^2\theta (\hat{p}_{\bar l})_x(\hat{p}_l)_{\bar{x}} +\beta^2 (\hat{p}_{\bar l})_y(\hat{p}_l)_{\bar{y}} \right. \nonumber \\
&& \left. + [\beta^2 +(2-\beta^2)\cos^2\theta](\hat{p}_{\bar l})_z(\hat{p}_l)_{\bar{z}} \right. \nonumber \\
&& \left. -\frac{2}{\gamma} \cos\theta \sin\theta ( (\hat{p}_{\bar l})_x(\hat{p}_l)_{\bar{z}}+(\hat{p}_{\bar l})_z(\hat{p}_l)_{\bar{x}} ) \right\}  / [(2-\beta^2)+\beta^2 \cos^2\theta]     
\end{eqnarray}
This angular distribution has been obtained by summing over all the annihilating quark-antiquark helicities. Clearly we want to separate those helicities to allow for transverse momenta polarized quark distributions. This will be done carefully for the gluon fusion mechanism.

The gluon fusion mechanism for {\bf unpolarized} gluons, is summed over gluon helicities. This gives rise to a higher order angular distribution 
due to the combination of two spin 1 gluons. 
%\bf{GLUON GLUON FUSION}
\begin{eqnarray}
W(\theta,p,p_{\bar l},p_l) & = & \frac{1}{4}-\frac{1}{4}
\left\{ [p^4\sin^4\theta+m^4](\hat{p}_{\bar l})_x(\hat{p}_l)_{\bar{x}} +
[p^2(p^2-2m^2) \sin^4\theta -m^4](\hat{p}_{\bar l})_y(\hat{p}_l)_{\bar{y}}\right. \nonumber\\
                           &   & \mbox{}  +[p^4 \sin^4\theta-2p^2(p^2-m^2)\sin^2\theta+m^2(2p^2-m^2)]
 (\hat{p}_{\bar l})_z(\hat{p}_l)_{\bar{z}} \nonumber\\                          
                           &   & \mbox{} + 2 m p^2\sqrt{p^2-m^2} \cos\theta
\sin^3\theta[(\hat{p}_{\bar l})_x(\hat{p}_l)_{\bar{z}} 
- (\hat{p}_{\bar l})_z(\hat{p}_l)_{\bar{x}} ]   \}   \nonumber\\
% + ([p^2-m^2]\
%                           &   & \mbox{} \left. +[p^2+m^2] cos^2\theta)
%(\hat{p}_{\bar l})_z(\hat{p}_l)_{\bar{z}}]
                          &   & \mbox{} / \left[ p^2(2m^2-p^2)\sin^4\theta+2p^2(p^2-m^2)\sin^2\theta +m^2(2p^2-m^2)\right] \\
%\right\}
%\nonumber
 =\frac{1}{4} &-& \frac{1}{4} \left\{ [ (1-\beta^2)^2+\sin^4\theta)] (\hat{p}_{\bar l})_x(\hat{p}_l)_{\bar{x}} \right. \nonumber \\
&& \left. \hspace{0.5in}  + [-(1-\beta^2)^2 -(1-2\beta^2) \sin^4\theta ] (\hat{p}_{\bar l})_y(\hat{p}_l)_{\bar{y}} 
 \right. \nonumber \\
 && \left.  \hspace{0.5in} +[(1-\beta^4) -2 \beta^2 \sin^2\theta +\sin^4\theta ] (\hat{p}_{\bar l})_z(\hat{p}_l)_{\bar{z}} \right. \nonumber \\
&& \left. \hspace{0.7in} +  2\frac{\beta}{\gamma} \sin^3\theta \cos\theta   [(\hat{p}_{\bar l})_x(\hat{p}_l)_{\bar{z}} 
- (\hat{p}_{\bar l})_z(\hat{p}_l)_{\bar{x}} ] \right\}  \nonumber \\
&& \hspace{0.7in}  / \left[ (1-\beta^4) +2\beta^2 \sin^2\theta+(1-2\beta^2) \sin^4\theta \right] 
\end{eqnarray}
where $m$ is the top quark mass, $\theta$ is the top quark production
angle in the quark-antiquark or ${\bar t} \,  t$ CM frame, $p$ is the light quark or gluon CM momentum, $\beta$ is the magnitude of the relativistic velocity of the top or antitop quark in the CM, 
$\hat{p}_{\bar{l}}$ is the $l^+$ momentum direction in the top rest frame
and $\hat{p}_l$ is the corresponding $l^-$ direction in the antitop rest
frame. For the $q \bar{q}$ case, a large opening angle between the leptons
is favored.
The weighting factor is combined with the probability distribution
function for the production of a top pair, each of mass m, $P_i(m)$~\cite{DalGol2}.
% an improved analysis can be performed.

It can be seen that the for  light quark-antiquark annihilation into heavy quark pairs (at tree
level in QCD), the spins of the heavy quarks tend to be aligned,
reflecting the helicity of the intermediate virtual gluon. Because
annihilation dominated over gluon fusion for Tevatron top
production,
%~\cite{dchang}, 
this alignment of spins was preferred. That effect
is diluted by the smaller contribution from gluon fusion. 
The reverse is true at the LHC.
%%%%%%%%%%%
%%%%%%%%%%%

By fixing the mass of the top quark and expressing the lepton directions
in an ``optimized basis''~\cite{parke}, the spin correlations can be
expressed in a simpler form than $W(\theta,p,p_{\bar l},p_l)$ above. Then
the amount of spin alignment is given by a single parameter for each
event, $\kappa$, which is near 1 for the Standard Model. The D0 group  
verified that the top pair spins tend to be correlated as predicted, with
their six events giving $\kappa > -0.25$, and confidence level of
68 \%~\cite{d0spin}. Similar measurements have now been performed at the LHC by the ATLAS group~\cite{ATLAS}.  Using the mass dependent form of the density matrix
above will enable experimenters to test the variation of the value
of $\kappa$ with different assumed masses, thereby connecting to the
uncorrelated analysis results.

We now separate the dilepton angular distributions into different components for the four different combinations of gluon distributions, shown in Table~\ref{dtable}. In particular we concentrate on the (LP.LP) case, which measures the linearly polarized gluon pair.
\begin{eqnarray}
W^{(LP,\, LP)}(\theta,p,p_{\bar l},p_l) & = &- \frac{1}{4}+\frac{1}{4} \left\{ [(1-\beta^4)+\beta^2 \sin^2\theta(-2+(2-\beta^2)\sin^2\theta)] (\hat{p}_{\bar l})_x(\hat{p}_l)_{\bar{x}} \right. \nonumber \\
&& \left. \hspace{0.5in}  +  [(1-\beta^4)+\beta^2 \sin^2\theta(2-\beta^2\sin^2\theta)] (\hat{p}_{\bar l})_y(\hat{p}_l)_{\bar{y}} 
 \right. \nonumber \\
 && \left.  \hspace{0.5in} +[-(1-\beta^2)^2 + \beta^2 (2-\beta^2)\sin^4\theta ] (\hat{p}_{\bar l})_z(\hat{p}_l)_{\bar{z}} \right. \nonumber \\
&& \left. \hspace{0.7in} -4\frac{\beta^2}{\gamma} \sin^3\theta \cos\theta   [(\hat{p}_{\bar l})_x(\hat{p}_l)_{\bar{z}} 
- (\hat{p}_{\bar l})_z(\hat{p}_l)_{\bar{x}} ] \right\}  \nonumber \\
&& \hspace{0.7in}  / \left[ (1-\beta^2)^2 +\beta^4 \sin^4\theta \right] 
\end{eqnarray}
In Figure \ref{weights} we show the directional correlation distributions for an unpolarized gluon distribution and a linear-transverse polarized gluon distribution. We have not included any particular values of gluon distribution functions. For that, we would convolute our spectator model distributions with the weights. The distributions are rich in dependences on the energies and angles for the $t+\bar{t}$ pair and the dilepton momenta. The $W_{i\, j}$ is the $\theta$ and $\beta$ dependent factor multiplying the Cartesian components of $\hat{p}(\mu^+)_i, \; \hat{p}(\mu^-)_j$, plotted for {\bf varying $\beta$}, the magnitude of relativistic velocity of the top in the $t+\bar{t}$ Center-of-Mass frame. The lepton momenta are determined in their top or antitop rest frames. Coordinates in the $t+\bar{t}$ CM are determined as follows. The $t+\bar{t}$ pair have momentum $\vec{p}_{t+\bar{t}}$ in the p+p collider CM. In the $t+\bar{t}$ CM the pair of gluons (or quark-antiquark) has zero 3-momentum, so the orientation of one gluon relative to the top in the $t+\bar{t}$ CM is the angle $\theta$ boosted from the p+p CM. 

For illustration we chose the polar angle of the $t+\bar{t}$ CM to be $\theta=\pi/8$ and varied $\beta$. The resulting weighting factors are remarkable for the clear distinction between unpolarized and polarized gluons. 

It is important to note that linearly polarized gluons have to carry transverse momenta, $k_T$, relative to the nucleon direction (in order to have an azimuth fixed). So in the lab frame the pair of gluons, with $\vec{k}_{T1}, \, \vec{k}_{T2}$,  has transverse momentum $\vec{k}_{T1} + \vec{k}_{T2}$ that will be imparted to the $t+\bar{t}$. With $k_1+k_2=p_t + p_{\bar{t}}$ and the invariants $(p_t - k_1)^2$, along with the boost from the lab to the $t+\bar{t}$ CM, and $\beta$, the transverse momenta are determined. Asymmetries for the lepton momenta can be determined to take advantage of the pronounced separation between polarized and unpolarized gluons.

Work in progress will present the full $t+\bar{t}$ angular distributions expected when the model calculations for the gluon distributions are included.

\acknowledgements
Work on gluon distributions was done with the collaboration of J.O. Gonzalez Hernandez and J. Poage. Some of the content was presented in J. Poage's Tufts University Ph.D. dissertation (2017).
The interest of Krzysztof Sliwa in this work is appreciated. We thank experimental colleagues for useful comments. 
We are grateful to the organizers of DPF2017 for a productive meeting. 
Work of S.L. supported by U.S. D.O.E. grant DE-SC0016286.

\end{document}